\documentclass[10pt, conference]{IEEEtran}
\IEEEoverridecommandlockouts
\usepackage{cite}
\usepackage{amsmath,amssymb,amsfonts}
\usepackage{algorithmic}
\usepackage{graphicx}
\usepackage{textcomp}
\usepackage{xcolor}
\usepackage{amsmath}
\usepackage{nccmath}
\usepackage{amssymb}
\usepackage{amsmath}
\usepackage{xfrac}
\usepackage{leftidx}
\usepackage{cite}
\usepackage{dblfloatfix}
\usepackage{algorithm}
\usepackage{algorithmic}
\usepackage{braket}
\usepackage{physics}
\usepackage{soul}
\usepackage{stackengine}
\usepackage{gensymb}
\usepackage{svg}
\usepackage{multicol}
\usepackage[hidelinks,colorlinks=true,linkcolor=black,citecolor=black]{hyperref}
\def\BibTeX{{\rm B\kern-.05em{\sc i\kern-.025em b}\kern-.08em
    T\kern-.1667em\lower.7ex\hbox{E}\kern-.125emX}}
\begin{document}

\title{Distillation-Enhanced Continuous-Variable Quantum Teleportation for Satellite Communication Networks\\
}

\author{\IEEEauthorblockN{1\textsuperscript{st} Lia Suci Waliani}
\IEEEauthorblockA{\textit{Department of Electrical Engineering} \\
\textit{École de Technologie Supérieure, University of Québec}\\
Montreal, Canada \\
lia-suci.waliani.1@ens.etsmtl.ca}
\and 
\IEEEauthorblockN{2\textsuperscript{nd} Georges Kaddoum}
\IEEEauthorblockA{\textit{Department of Electrical Engineering} \\
\textit{École de Technologie Supérieure, University of Québec}\\
Montreal, Canada \\
georges.kaddoum@etsmtl.ca}
\and
\IEEEauthorblockN{\hspace{1cm} 3\textsuperscript{rd} Mahdi Chehimi}
\IEEEauthorblockA{\hspace{1.2cm}\textit{Department of Electrical and Computer Engineering } \\
\hspace{1.2cm} \textit{American University of Beirut}\\
\hspace{1.2cm} Beirut, Lebanon \\
\hspace{1.2cm} \textit{Department of Electrical and Computer Engineering} \\
\hspace{1.2cm}\textit{Western University}\\
\hspace{1.2cm}London, Ontario, Canada \\
\hspace{1.2cm}mc127@aub.edu.lb}
\and
\IEEEauthorblockN{\hspace{1cm}4\textsuperscript{th} Shahan Hawatian}
\IEEEauthorblockA{\hspace{1cm}\textit{Department of Electrical and Computer Engineering} \\
\hspace{1cm}\textit{American University of Beirut}\\
\hspace{1cm}Beirut, Lebanon \\
\hspace{1cm}sh11@aub.edu.lb }
}

\maketitle

\begin{abstract}
Quantum teleportation (QT) over satellite-based free-space optical (FSO) channels is a promising approach for long-distance quantum communication. However, its performance is significantly degraded by atmospheric loss and turbulence. In this paper, we investigate continuous-variable (CV) QT in a dual-downlink scenario, where a satellite distributes entangled states to two ground stations. To mitigate channel-induced degradation, we employ a non-Gaussian entanglement distillation protocol based on the sequential application of photon addition and photon subtraction (PA–PS) on the weaker channel. The results show that the proposed scheme improves teleportation fidelity by up to $7.7\%$ and enhances entanglement negativity by approximately $105\%$ in the low-to-moderate (below $600$~km) loss regime. In addition, we identify an optimal squeezing parameter that balances entanglement strength and noise sensitivity. Taken together, these results demonstrate the effectiveness of PA–PS distillation for improving CV quantum communication in realistic satellite networks. We further characterize the trade-off between fidelity gain and the heralded success probability of the protocol.
\end{abstract}


\section{Introduction}
\IEEEPARstart{Q}{uantum} teleportation (QT) was first proposed by Bennett \textit{et al.} in 1993 \cite{bennett1993teleporting} to address a challenge in quantum information transfer. Specifically, the challenge was that an unknown quantum state could not be reliably transmitted or copied using a classical or direct method. QT enables the transfer of an unknown quantum state between two distant parties without physically transmitting the particle carrying the state. Instead, QT relies on shared entanglement and classical communication between two communicating parties known as Alice and Bob \cite{cacciapuoti2020QT,chehimi2022physics}. 

Shared entanglement, typically realized through the Einstein–Podolsky–Rosen (EPR) pair states, plays a fundamental role in the QT protocol. One approach to generate entanglement is through continuous-variable (CV) systems, where entangled states are produced using squeezed light generated from nonlinear optical processes \cite{CV_QC}. In this scheme, two squeezed optical modes are combined at a beam splitter to generate a two-mode squeezed vacuum (TMSV) state. This state is then transmitted through a quantum channel. In the receiver, quadrature variables are typically measured using high-speed and high-efficiency optical devices, such as homodyne or heterodyne detectors \cite{ shi2023continuous}. This process is referred to as the entanglement distribution phase.

The entanglement distribution phase can be realized using fiber optics and free-space optics (FSO) channels. However, in regions with obstacles or limited infrastructure, the deployment of fiber-optic networks can be challenging \cite{nguyen2023enhancing}. Moreover, fiber-optic channels suffer from a significant photon loss over long transmission distances \cite{fiber_bipartite}. In such scenarios, the FSO quantum channel offers a promising solution to support the limitations of the fiber-optic channel. In an FSO channel, quantum information is wirelessly transmitted through the air between two remote nodes. These FSO links can be implemented either as terrestrial networks or as space-based FSO channels by involving satellites \cite{survey_cv_satellite}. The use of space-based FSO channels,
which support terrestrial networks in remote areas, enables global-scale quantum communication in the near future.

However, despite the capability of the FSO channel to support terrestrial networks, several challenges persist. One of the main limitations arises from channel losses and atmospheric turbulence in FSO links \cite{chehimi2025reconfigurable}. As the quantum signal propagates from the satellite to ground stations (GSs), the entanglement will be degraded due to atmospheric effects. To date, many previous studies demonstrated feasibility of satellite networks, particularly in the CV regime \cite{pirandola_cv-sat, uplink_cv}. In \cite{pirandola_cv-sat}, the authors analyzed the effects of atmospheric turbulence on the CV entanglement distribution and quantum teleportation in the optical regime between a GS and a satellite. Their results revealed that downlink communication is feasible for a low Earth orbit (LEO) satellite. However, for the uplink scenario, an intermediate station is required to facilitate communication between the satellite and the GS. In another relevant study \cite{uplink_cv}, the authors proposed a practical diversity-assisted scheme for the distribution of uplink entanglement and the transfer of coherent states and analyzed the evolution of the quantum state during these processes. However, the aforementioned study did not address the entanglement degradation: namely, because of the channel loss and atmospheric turbulence, the entangled pairs distributed between the satellite and the GS are severely degraded. Therefore, it was argued that additional protocols are required to recover high-quality entanglement from degraded entangled states \cite{lee2013distillation}.

To address fidelity degradation, conventional Gaussian operations alone cannot increase fidelity of the entanglement, which motivates the use of non-Gaussian distillation techniques such as noiseless linear amplification (NLA) \cite{xiang2010NLA}, quantum scissors (QS) \cite{xia2025_QS}, 
photon subtraction (PS) \cite{zhang2010photonsubstraction, wang2024PSswapping}, and photon addition (PA) \cite{lee2011psa}. However, under high channel loss, the quantum scissor truncates significant information from quantum state. Therefore, PS and PA emerge as a promising solution for satellite communication networks (SCNs). 
In \cite{villasenor2021distillnon}, the authors analyzed non-Gaussian operations to enhance the entanglement of a TMSV state, showing that a combination of PA and PS can enhance entanglement in terms of both fidelity and negativity. However, their analysis has several limitations: the channel is modeled as a generic noisy channel without accounting for the physical origin of the loss (diffraction, atmospheric attenuation, turbulence, and pointing errors), the study is restricted to a single-link configuration, and no geometric or environmental dependence is investigated. Moreover, the interplay between distillation and channel-specific effects such as turbulence strength, zenith angle, and beam wandering remains unexplored.

Motivated by these gaps, the contributions of this paper are as follows:
\begin{itemize}
    \item We develop a dual-downlink CV-QT framework over a physical satellite FSO channel that jointly incorporates diffraction, atmospheric attenuation, Hufnagel-Valley turbulence, beam wandering, pointing errors, and detector inefficiency.
    \item We propose a selective PA-PS distillation policy that applies the non-Gaussian operation only to the weaker downlink, preserving the overall success probability while maximizing entanglement quality.
    \item We provide a systematic analysis of how distillation gains depend on physical channel parameters, such as zenith angle and turbulence strength. This has not been addressed in prior non-Gaussian distillation studies.
    \item We identify an optimal squeezing parameter that balances entanglement strength against loss-induced noise amplification, and we characterize the fidelity and success probability trade-off of the protocol.
\end{itemize}

\textit{Organization}: The remainder of this paper is organized as follows. Section \ref{Entanglement distribution} describes the entanglement distribution phase. In Section \ref{QT and distillation}, we introduce the proposed PA-PS distillation protocol. Section \ref{numerical_result} presents the numerical result. Section \ref{conclusion} concludes this paper. 

\textit{Notation}:
$(\cdot)^{-1}$ indicates inversion.
$(\cdot)^{*}$ indicates a complex conjugate.
$|\cdot|$ refers to the absolute.
$\hat{a} $ and $ \hat{a}^{\dagger}$ refers to annihilation and creation operators, respectively. 
$\mathbb{I}$ denotes an identity matrix.

\section{Entanglement Distribution Phase}\label{Entanglement distribution}

In this paper, we consider a dual-downlink scenario where the satellite acts as a photon source, generating entangled photon pairs and distributing the two modes to ground stations A and B (see Fig. \ref{fig: dd}). The entangled state is described as a TMSV state, which is a fundamental entangled Gaussian state in CV quantum systems. The TMSV is generated by applying a two-mode squeezing operator to the vacuum states\footnote{The vacuum states represent the quantum state of a single electromagnetic mode when no photons are present. In the phase-space representation, the vacuum state is centered at the origin, with mean quadratures $\langle\hat{q}\rangle=0$, $\langle\hat{p}\rangle=0$, and a covariance matrix that is equal to the identity under $\hbar=2$ normalization, i.e., $\mathbf{V}_{\text{vac}}=\mathbb{I}$.} of two independent bosonic modes labeled as modes A and B. The two-mode squeezing operator is given by \cite{CV_book}:
\begin{figure}[tb]
    \centering
    \includegraphics[width=1\linewidth]{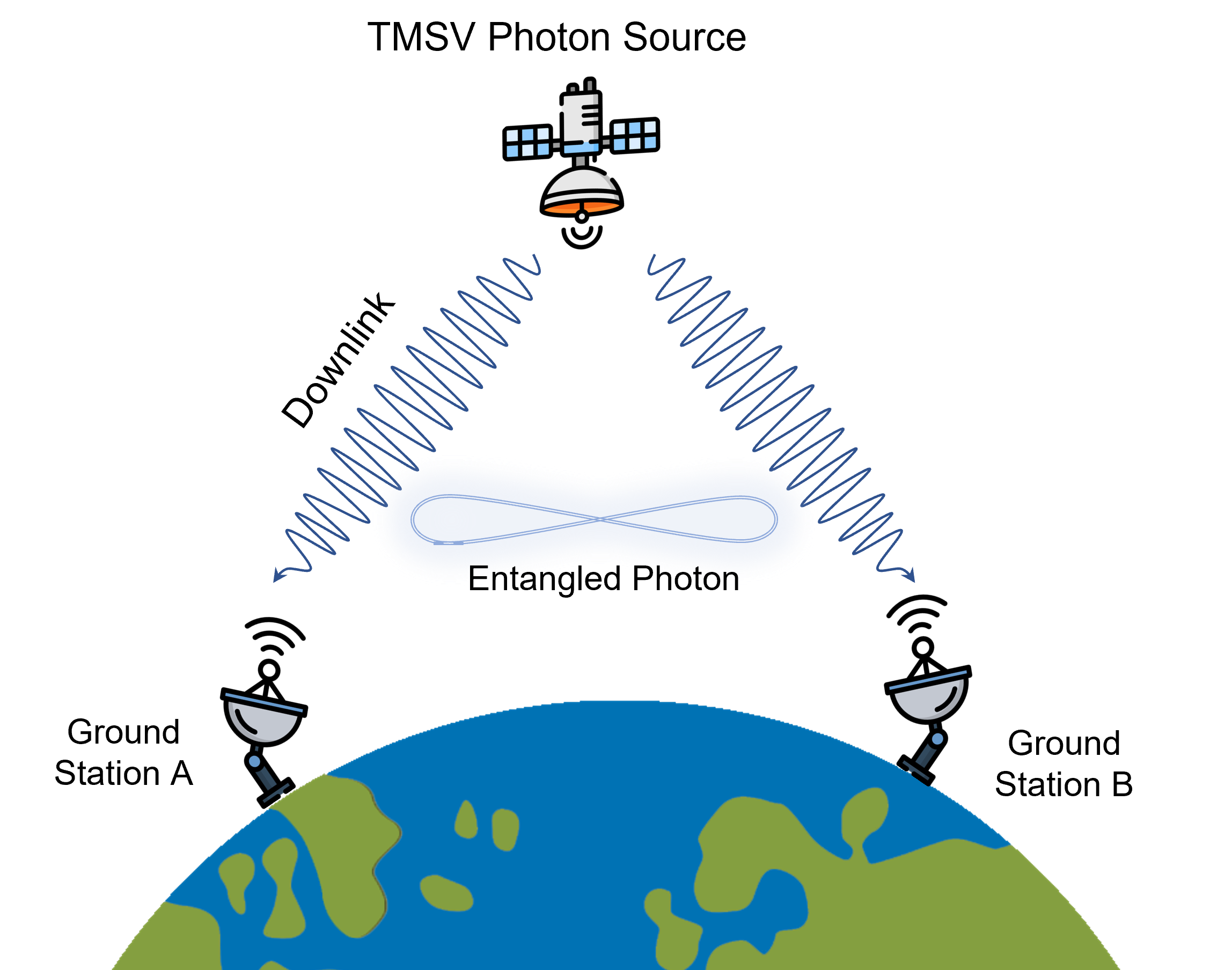}
    \caption{Dual-downlink FSO quantum link where a satellite generates a TMSV entangled photon pair. Each mode is sent to two GSs through independent channels with different losses.}
    \label{fig: dd}
\end{figure}

\begin{figure*}[tb]
    \centering
    \includegraphics[width=1\linewidth]{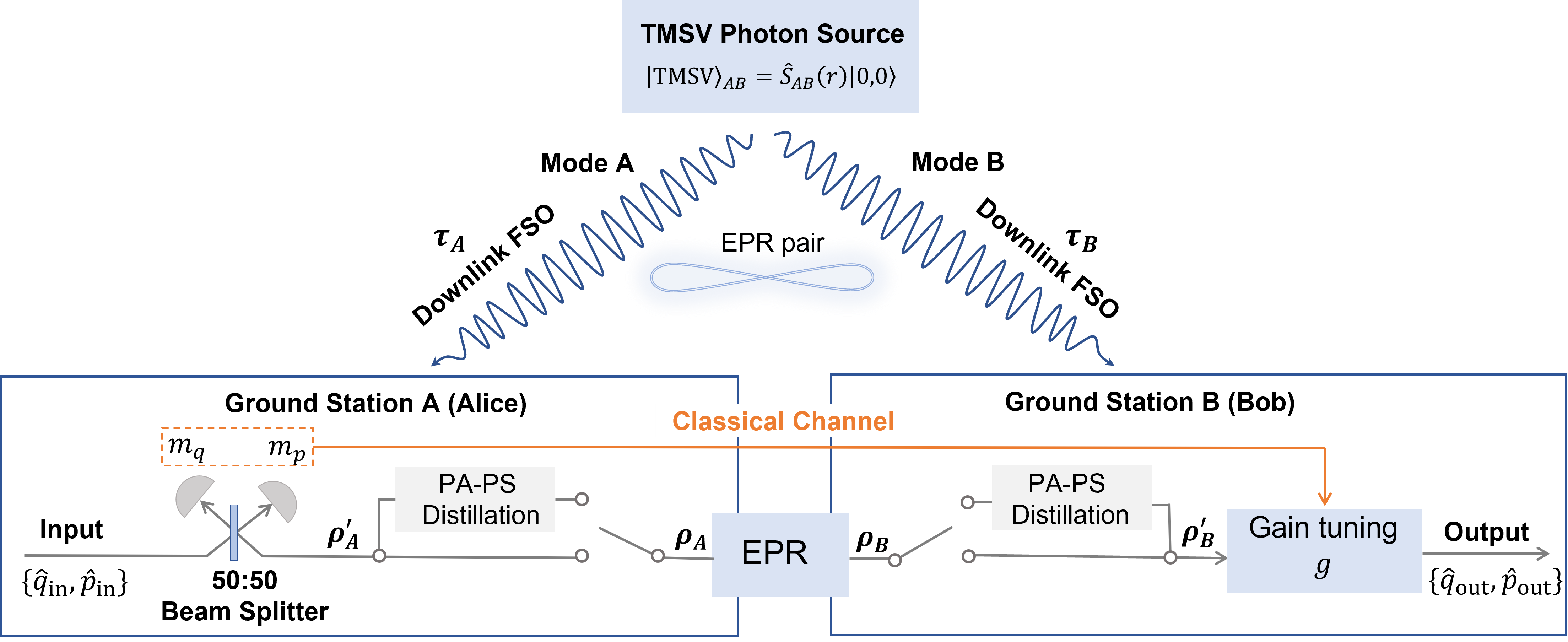}
    \caption{Dual-downlink CV-QT over satellite-based FSO links, where a satellite distributes a TMSV entangled pair to two ground stations through asymmetric channels. A switch selects the weaker link based on its transmissivity and applies PA–PS distillation to enhance entanglement.}
    \label{fig: system model}
\end{figure*}
\begin{equation}
     \hat{S}_{AB}(r) = \exp\left[r \left(\hat{a}_1\hat{a}_2 - \hat{a}_1^{\dagger}\hat{a}_2^{\dagger}\right)\right],
\end{equation}
where $r \in \mathbb{R}$ is the squeezing parameter and $\hat{a}^{\dagger}$ is the creation operator. Therefore, the TMSV state can be written as follows:
\begin{equation}
    \ket{\text{TMSV}}_{A,B} = \hat{S}_{AB}(r)\ket{0,0}.
    \end{equation}

In the phase-space representation, the TMSV state has zero mean quadratures, and its covariance matrix is given by Eq. \eqref{m_tmsv}:
    \begin{equation} \label{m_tmsv}
    \mathbf{V}_{\mathrm{TMSV}} =
    \begin{pmatrix}
    \alpha\mathbb{I} & \gamma\mathbf{Z} \\
    \gamma\mathbf{Z} & \beta\mathbb{I}
    \end{pmatrix},
\end{equation}
where $\mathbb{I}$ denotes the identity matrix and $\mathbf{Z} = \operatorname{diag}(1, -1)$. The parameters are defined as $\alpha=\cosh({2r})$, $\beta=\cosh({2r})$, and $\gamma=\sinh(2r)$. Equivalently, the covariance matrix can be written as follows:
 \begin{equation} 
    \mathbf{V}_{\mathrm{TMSV}} =
    \begin{pmatrix}
    \cosh(2r)\mathbb{I} & \sinh(2r)\mathbf{Z} \\
    \sinh(2r)\mathbf{Z} & \cosh(2r)\mathbb{I}
    \end{pmatrix}.
\end{equation}

Subsequently, the TMSV state is transmitted to the GS through an FSO channel. During transmission, the propagating quantum signal is subject to several degradation mechanisms that reduce its quality before reaching the ground station. These include diffraction-related transmissivity loss, atmospheric attenuation, atmospheric turbulence, beam broadening, beam wandering, and detector inefficiency, all of which jointly affect transmissivity and noise characteristics of the received quantum state.

Let $z$ be the propagation distance between the ground station and the satellite. The vertical separation between the satellite and the ground station is given by $\Delta\mathcal{H} = \mathcal{H}-\mathcal{H}_0$, where $\mathcal{H}$ is the altitude of the satellite, while $\mathcal{H}_0$ is the altitude of the ground station. For a slant path with a zenith angle $\theta$, the propagation distance $z$ can be expressed as follows:
\begin{equation}
    z=\sqrt{(\Delta \mathcal{H})^2 + 2 \Delta \mathcal{H} \mathcal{R} + \mathcal{R}^2 \cos^2 \theta} -\mathcal{R} \cos \theta,
\end{equation}
where $\mathcal{R}$ denotes the radius of the Earth.

First, we consider the diffraction-induced transmissivity as a source of degradation. This occurs because the optical beam spreads during propagation, and only a fraction of the transmitted field is collected by the receiver aperture. This aperture-capture effect is modeled through a diffraction-related transmissivity, denoted by $\tau_{\mathrm{diff}}$, which can be defined as follows:
\begin{equation}
    \tau_{\text{diff}} = 1 - \exp\left[ - \frac{2\mathcal{A}^{2}}{\varpi_0^2 \bigg[\left(1-\frac{z}{\mathcal{R}_0} \right)^2 + \left(\frac{z}{z_\mathcal{R}} \right)^2 \bigg]} \right],
\end{equation}
where ${\mathcal{A}}$ denotes the receiver aperture radius, $\varpi_0$ is the initial beam waist, and $z_{\mathcal{R}} = \pi \frac{\varpi_{0}^{2}}{\lambda}$ represents the Rayleigh range, with $\lambda$ being the wavelength of the optical signal.

Next, we consider atmospheric attenuation, or the exponential loss of optical signals caused by absorption and scattering. Absorption occurs when photons are absorbed by atmospheric gases, whereas scattering occurs when photons are deflected away from the receiver aperture as they propagate through the atmosphere. The atmospheric attenuation, denoted by  $\tau_{\mathrm{atm}} \in (0, 1]$, can be modeled as an exponential loss along the propagation path and is given by the following \cite{pirandola_cv-sat}:
\begin{equation}
    \tau_{\text{atm}} = \exp \big[{-\mathfrak{a}_0 g(\mathcal{H},\theta)}\big].
\end{equation}
Here, $\mathfrak{a}_0$ denotes an extinction coefficient at sea level and $g(\mathcal{H},\theta)$ integrates the atmospheric density profile along the propagation path. The function $g(\mathcal{H},\theta)$ is given by the following:
\begin{equation}
    g(\mathcal{H},\theta) = \int_{0}^{z(\mathcal{H},\theta)} dy \, e^{-h(y,\theta)/\bar{h}}.
\end{equation}

Moreover, atmospheric turbulence refers to random fluctuations in the air's refractive index along a propagation path caused by variations in temperature and pressure within the atmosphere. In weak-turbulence, these effects can be treated as perturbations to Gaussian beam propagation and are fully characterized by the spherical-wave coherence length $\rho$, which can be expressed as $\rho_0 = [1.46k^2I_0(z)]^{\frac{-3}{5}}$, where $I_0(z)$ denotes the path-integrated turbulence strength for a spherical optical wave propagating over a distance $z$, which can be calculated as follows \cite{pirandola_cv-sat}:
\begin{equation}
    I_0(z) = \int_{0}^{z} d\xi \left( 1 - \frac{\xi}{z} \right)^{5/3}
C_n^2\!\bigl(\mathcal{H}(\xi,\theta)\bigr),
\end{equation}
where $z \lesssim k \, \min\{ 2 \mathcal{A}, \rho_0 \}^2$ denotes the weak-turbulence condition for beam with wavenumber $k = \frac{2\pi}{\lambda}$. Quantity $C_n^2$ denotes the refractive-index structure constant, which characterizes the strength of atmospheric turbulence. It can be modeled using the Hufnagel-Valley (HV) atmospheric turbulence model, which can be expressed as follows \cite{hufnagel1964modulation}:
\begin{equation}
\begin{aligned}
    C_n^2 =\;& \underbrace{5.94 \times 10^{-53} \left(\frac{v}{27}\right)^2 \mathcal{H}^{10} e^{-\mathcal{H}/1000}}_{\text{High-altitude turbulence}} \\
    &+ \underbrace{ 2.7 \times 10^{-16} e^{-\mathcal{H}/1500}}_{\text{Mid-altitude turbulence}} \\
    &+ \underbrace{A_{\mathrm{HV}} e^{-\mathcal{H}/100}}_{\text{Near-ground turbulence}},
\end{aligned}
\end{equation}
where $v$ denotes the root-mean-square (RMS) wind speed, while $A$ refers to turbulence strength associated with near-ground conditions, which can vary between daytime and nighttime. 

Turbulence causes the beam to broaden the diffraction limit and wander around the receiver axis. The deterministic loss due to diffraction and turbulence-induced beam spreading is described by short-term transmissivity $\tau_{\mathrm{st}}$, which can be expressed as follows:
\begin{equation}
    \tau_{\mathrm{st}} = 1 - \exp\left[-2\left(\frac{A}{\varpi_{\mathrm{st}}}\right)^2\right].
\end{equation}
\begin{figure}[tb]
    \centering
    \includegraphics[width=0.9\linewidth]{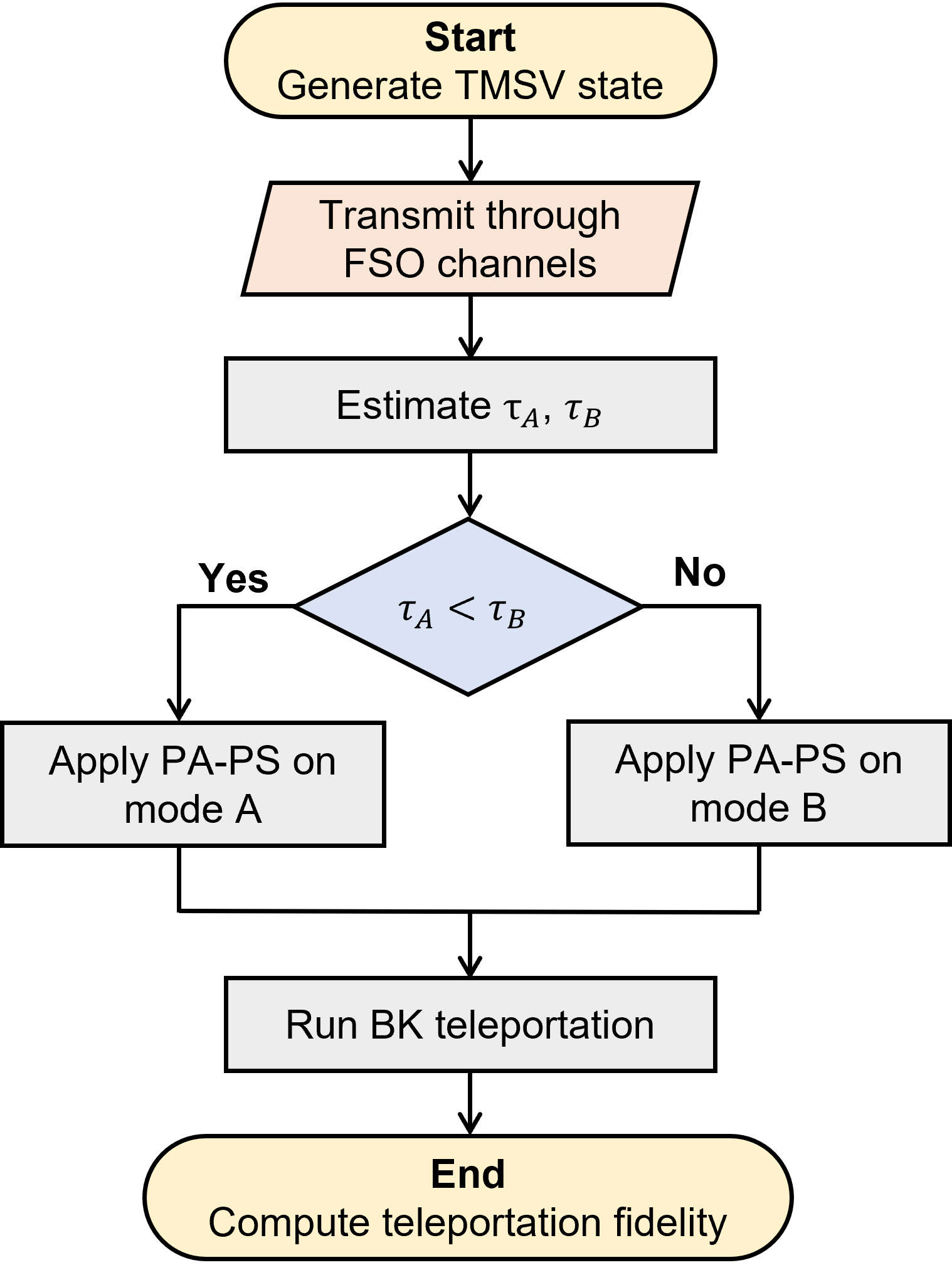}
    \caption{Flowchart of the selective distillation protocol for dual-downlink CV teleportation.}
    \label{fig: flowchart}
\end{figure}

Weak atmospheric turbulence increases the effective beam radius beyond the diffraction-limited value $\varpi_z$. This effect is taken into account by introducing a short-term beam waist $\varpi_{\mathrm{st}}$, which can be expressed as follows  \cite{pirandola_cv-sat}:
\begin{equation}
    \varpi_{\mathrm{st}}^{2} \simeq \varpi_{z}^{2} + 2 \left( \frac{\lambda{z}}{\pi \rho_{0}} \right)^{2} (1 - \phi)^{2},
\end{equation}
where $\phi = 0.33(\frac{\rho_0}{\varpi_0})^{1/3}$. 

The variance of turbulence-induced beam wandering is obtained from the difference between long-term and short-term beam waists, resulting in $\sigma_{\mathrm{TB}}$, which explicitly depends on the propagation distance, initial beam waist, and coherence length $\rho_0$. It can be expressed as follows:
 \begin{equation}
    \sigma_{\mathrm{TB}}^{2}
    = \varpi_{\mathrm{lt}}^{2}
    - \varpi_{\mathrm{st}}^{2}
    \simeq 0.1337 \lambda^{2}
    \frac{z^{2}}{\varpi_{0}^{1/3} \rho_{0}^{5/3}}.
\end{equation}
 Furthermore, when pointing errors are included, the total centroid variance $\sigma^2$ accounts for both the turbulence-induced motion and the mechanical misalignment effect, which can be expressed as follows:
\begin{equation}
    \sigma^2 = \sigma_{\mathrm{TB}}^2 + \sigma_{\mathrm{P}}^2,
\end{equation}
where $\sigma_{\mathrm{P}} = 10^{-6}z $ \cite{pirandola_cv-sat}.

By combining deterministic losses, the maximum transmissivity is given by the following:
\begin{equation}
\begin{aligned}
    \tau_{\mathrm{max}} = \tau_{\mathrm{st}} (\mathcal{Q}=0) \tau_{\mathrm{atm}} \tau_{\mathrm{eff}}.
\end{aligned}
\end{equation}

Since the beam displacement $\mathcal{Q}$ is a random variable, the overall channel transmissivity $\tau$ is also random. The effect of beam wandering is incorporated. The probability density function $P(\tau)$ induced by the distribution of $\mathcal{Q}$ can be expressed as follows \cite{pirandola_cv-sat}:
\begin{equation}
\begin{aligned}
P(\tau) = &
    \frac{\mathcal{Q}_0^{2}}{\gamma \sigma^{2}\,\tau}
    \left[
    \ln\!\left(\frac{\tau_{\max}}{\tau}\right)
    \right]^{\frac{2}{\gamma}-1} \\
    & \exp\!\left[
    -\frac{\mathcal{Q}_0^{2}}{2\sigma^{2}}
    \left(
    \ln\!\left(\frac{\tau_{\max}}{\tau}\right)
    \right)^{\frac{2}{\gamma}}
    \right],
    \end{aligned}
\end{equation}
together with the mapping 
\begin{equation}
    \mathcal{Q} = \mathcal{Q}_0 \left[ \ln\!\left(\frac{\tau_{\max}}{\tau}\right) \right]^{1/\gamma}.
\end{equation}

\section{Entanglement Distillation and Teleportation Protocol}\label{QT and distillation}

The general procedure of the selective distillation protocol, worst-mode selection, and conditional application of PA-PS are summarized in the flowchart (see Fig.~\ref{fig: flowchart}). 
Let the transmissivities of the two modes be denoted by $\tau_A$ and $\tau_B$, corresponding to the two downlink channels (see~Fig.~\ref{fig: system model}). We treat the downlink transmissivities as random variables and evaluate the system performance via Monte Carlo averaging. Specifically, $\tau_A$ and $\tau_B$ are independently sampled from their respective channel distributions for each realization. This independence assumption is justified by the fact that the two downlinks propagate through different atmospheric columns toward distinct GSs and experience independent turbulence-induced fluctuations.

\begin{figure}[tb]
    \centering
    \includegraphics[width=0.9 \linewidth]{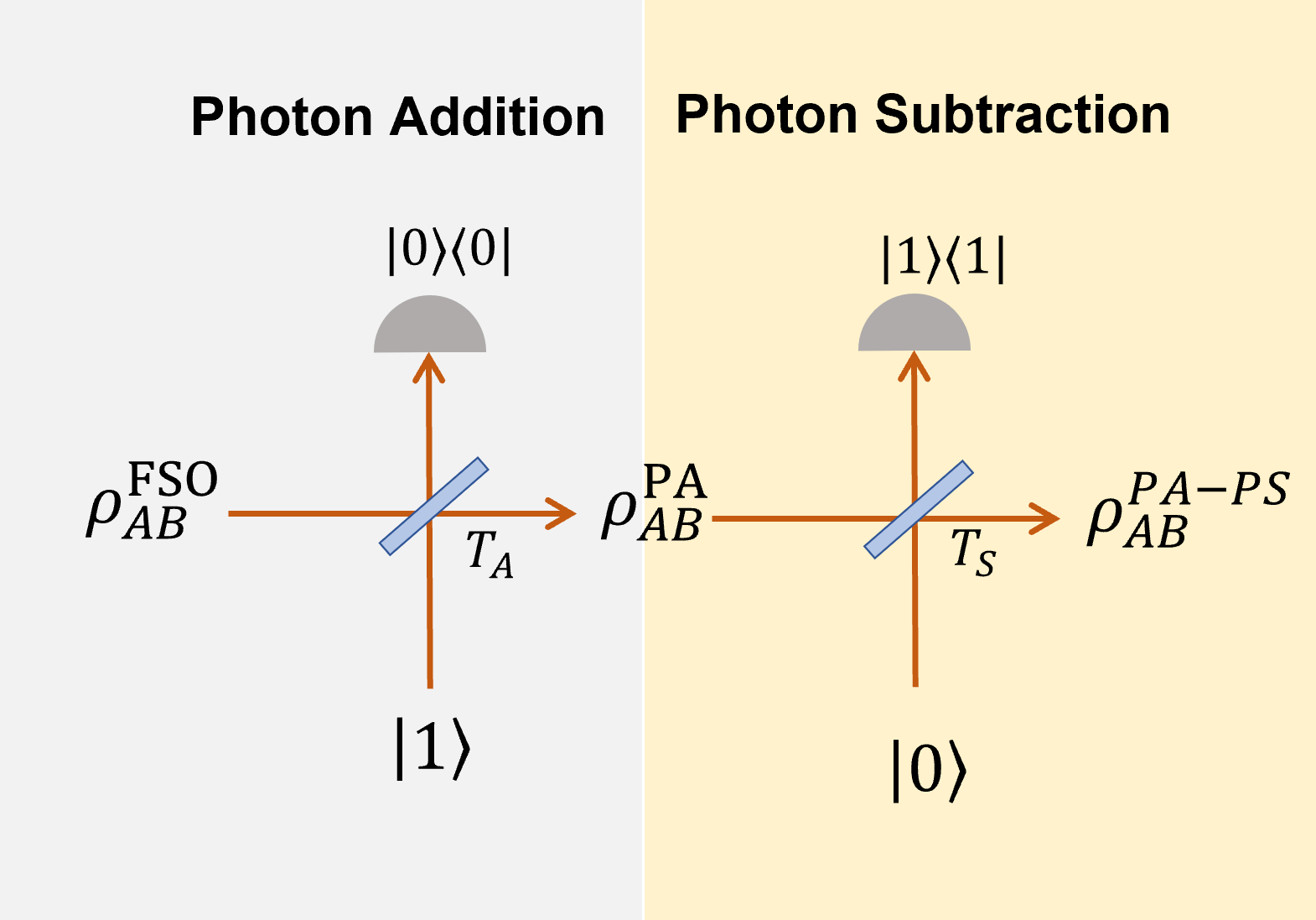}
    \caption{The PA-PS acting as the entanglement distillation protocol.}
    \label{fig: ldqs}
\end{figure}

After propagation through the dual downlink FSO channels, the two GSs receive a noisy entangled state. The corresponding characteristic function (CF) of the shared entanglement resource can be written as follows:
\begin{equation}
\begin{aligned}
    \chi_{\mathrm{ch}}(\zeta_A, \zeta_B)
    =& \exp\bigg[
    - \frac{1}{2}
    \big(
    \alpha_{\mathrm{ch}} |\zeta_A|^2
    + \beta_{\mathrm{ch}} |\zeta_B|^2
     \\&+ \gamma_{\mathrm{ch}} \zeta_A \zeta_B
    + \gamma_{\mathrm{ch}}^* \zeta_A^* \zeta_B^*
    \big)
    \bigg],
\end{aligned}
\end{equation}
where $\zeta_A$ and $\zeta_B$ are complex phase-space variables associated with modes A and B, respectively. Parameters $\alpha_{\mathrm{ch}} = \tau_A \alpha + (1-\tau_A)m_A$ and $\beta_{\mathrm{ch}} = \tau_B \alpha + (1-\tau_B)m_B$ denote the effective quadrature variances of mode A, where $m_A$ is the thermal noise in mode A and B, respectively, after channel transmission. Here, $m_A$ and $m_B$ represent the thermal noise contributions in modes A and B, respectively. In this paper, thermal noise is assumed to be negligible, such that $m_A \approx m_B \approx 1$. The term $\gamma_{\mathrm{ch}} =\sqrt{\tau_A \tau_B}\gamma$ represents the reduced correlation between modes.

In principle, the distillation protocol can be applied at both GSs. However, since the distillation process is probabilistic, applying it to both modes would significantly reduce the overall success probability. Each GS estimates its local channel transmissivity and communicates this information through a classical link.  The worst mode is then identified by selecting the minimum transmissivity, which can be described as follows:
\begin{equation}
    w = \arg \min \{{\tau_A(t), \tau_B(t)}\},
\end{equation}
where $w \in \{ A, B\}$ denotes the mode with the lowest transmissivity. The noisy two-mode state after propagation through the channels is described by the CF $f(\zeta_A, \zeta_B) $, which incorporates the effects of noise in both modes. 
To enhance shared entanglement, the distillation operation is then applied to the mode with the lowest transmissivity. 

In this study, we modeled the distillation protocol as a sequential combination of PA and PS (see Fig.~\ref{fig: ldqs}). To describe the distillation protocol, we begin with the PA first. Physically, this is a heralded process that conditionally adds one photon to the selected mode. Mathematically, photon addition corresponds to creation operator $\hat{a}^{\dagger}$ of a quantum state. If the initial state is $\ket{\psi}$, then after adding the photon, the state can be defined as follows:
\begin{equation}
    \ket{\psi_{\mathrm{add}}^k} \propto (\hat{a}^{\dagger})^k \ket{\psi},
\end{equation}
where $k$ denotes the number of photons. In the density operator representation, photon addition in a mixed state $\rho$ is described as follows:
\begin{equation}
    \rho_{\mathrm{out}} \rightarrow \hat{a}^\dagger \rho_{\mathrm{in}} \hat{a},
\end{equation}
up to normalization, where the normalization factor corresponds to the success probability of the heralded process. Additionally, the unnormalized CF can be written as follows:
\begin{equation}
\begin{aligned}
    \chi'_{\mathrm{PA}}(\zeta_A, \zeta_B)
   =& (T_{\mathrm{S}} - 1)\,
    \exp\!\left(\frac{|\zeta_w|^2}{2}\right)
    \frac{\partial^2}{\partial \zeta_w \partial \zeta_w^*}
    \\& \left[
    \exp\!\left(-\frac{|\zeta_w|^2}{2}\right)
    \, f(\zeta_A, \zeta_B, \sqrt{T_{\mathrm{S}}})
    \right],
\end{aligned}
\end{equation}
where $w \in \{ A, B\}$ denotes the mode in which the distillation is applied. $\zeta_A$ and $\zeta_B$ denote the complex phase-space variables associated with modes A and B, respectively. $T_{\mathrm{S}}$ is a parameter related to the beam splitter transmissivity.

Afterwards, PS is applied to the same mode. PS is a non-Gaussian, probabilistic operation where one or more photons are removed from an optical mode. Mathematically, it is modeled by annihilation operator $\hat{a}$. Therefore, the state can be defined as follows:
\begin{equation}
    \ket{\psi_{\mathrm{sub}}} \propto \hat{a} \ket{\psi}.
\end{equation}
In the density operator representation, photon subtraction in a mixed state $\rho$ is described as follows:
\begin{equation}
   \rho_{\mathrm{out}} \rightarrow \hat{a} \rho_{\mathrm{in}} \hat{a}^\dagger, 
\end{equation}
up to normalization. The unnormalized CF can be written as follows:
\begin{equation}
    \begin{aligned}
        \chi'_{\mathrm{PS}}(\zeta_A, \zeta_B) =& \frac{T_{\mathrm{S}} - 1}{T_{\mathrm{S}}}
        \exp\!\left(-\frac{|\zeta_w|^2}{2}\right)
        \frac{\partial^2}{\partial \zeta_w \partial \zeta_w^*} \\&
        \left[
        \exp\!\left(\frac{|\zeta_w|^2}{2}\right)
        \, f(\zeta_A, \zeta_B, \sqrt{T_{\mathrm{S}}})
        \right].
    \end{aligned}
\end{equation}
Finally, the full unnormalized CF PA-PS sequence is defined as follows  \cite{villasenor2021distillnon}:
\begin{equation} \label{Eq: paps a}
    \begin{aligned}
        \chi'_{\mathrm{PA-PS}}(\zeta_A, \zeta_B)
     = & (T_{\mathrm{S}} - 1)^2 \exp\left(-\frac{|\zeta_w|^2}{2}\right)
    \frac{\partial^2}{\partial \zeta_w \partial \zeta_w^*} \\&
    \bigg\{
    \exp\left(|\zeta_w|^2\right)
    \times
    \frac{\partial^2}{\partial \zeta_w \partial \zeta_w^*}
    \bigg[
    \exp\left(-\frac{|\zeta_w|^2}{2}\right) \\&
    f(\zeta_A, \zeta_B, T_{\mathrm{S}})
    \bigg]
    \bigg\},
    \end{aligned}
\end{equation} 
where the second-order differential operator $ \frac{\partial^2}{\partial \zeta_w \partial \zeta_w^*}$ represents the action PA-PS operation in phase space, while the exponential factor arises from the ordering transformation of the CF. The intermediate is defined as shown by the following equation:
\begin{equation}
    f(\zeta_A, \zeta_B, T_{\mathrm{S}}) = \int d^2 \xi \chi_{\mathrm{ch}} (\zeta_A, \zeta) \times \mathrm{K} (\zeta, \zeta_B, T_{\mathrm{S}}),
\end{equation}
where $\chi_{\mathrm{ch}} (\zeta_A, \zeta)$ denotes the CF of the state after transmission through the noisy channels, while $\mathrm{K} (\zeta, \zeta_B, T_s)$ is the kernel function that describes the beam splitter interaction in the photon subtraction stage.

\begin{table}[tb]
    \centering
    \caption{Simulation Parameters}
    \resizebox{\columnwidth}{!}{
    \begin{tabular}{|c|c|c|}
    \hline
         \textbf{Parameter}& \textbf{Definition} & \textbf{Value} \\ \hline \hline
         $\mathcal{H}_0 $ & GS altitude & $0$~m \\ \hline 
         $\lambda$ & wavelength & $800$~nm \\ \hline
         $\bar{h}$ & atmospheric scale height & $6600$~m \\ \hline
         $v$ & root-mean-square wind speed & $21$~m/s \\ \hline
         $\mathcal{R}$ & Earth's radius & $6.371\times10^6$~m \\ \hline
         $\mathcal{A}_A$ & receiver aperture radius for link~$A$  & 0.7~m \\ \hline
         $\mathcal{A}_B$ & receiver aperture radius for link~$B$ & 0.6~m \\ \hline
         $\varpi_{0,A}$ & initial beam waist for link~$A$ & 0.2~m \\ \hline
         $\varpi_{0,B}$ & initial beam waist for link~$B$ & 0.28~m \\ \hline
         $A_{\mathrm{HV}}$ & HV turbulence constant & $ 2.75 \times 10^{-14} \, \text{m}^{-2/3}$ \\ \hline
         $\tau_{\mathrm{eff}}$ & detector efficiency & 1 \\ \hline
         $r$ & default squeezing parameter & 0.6 \\ \hline
         $g$ & gain & 1 \\ \hline
         $\eta^2$ & homodyne efficiency & 0.49 \\ \hline
         $\mathfrak{a}_0$ & extinction coefficient & $2.5 \times{10}^{-6}~{\mathrm{m}}^{-1}$  \\ \hline
         $T_S$ & Beam-splitter transmissivity & 0.95 \\ \hline
         $N_{\mathrm{cutoff}}$ & Fock space cut off & 8 \\ \hline
         $N_{\mathrm{MC}}$ & Monte Carlo trials & $10^3$ \\ \hline
    \end{tabular}
    }
    \label{tab: param}
\end{table}

Afterwards, the normalized CF is obtained by dividing the unnormalized CF by its value at zero:
\begin{equation}
    \chi_{\mathrm{dis}} (\zeta_A, \zeta_B) = \frac{\chi'_{\mathrm{PA-PS}}(\zeta_A, \zeta_B)}{\chi'_{\mathrm{PA-PS}}(0, 0)}.
\end{equation}
Denominator $\chi'_{\mathrm{PA-PS}}(0, 0)$ is the success probability of the heralded PA-PS operation. In other words,~$\chi'_{\mathrm{PA-PS}}(0, 0) = p_{succ}$. This quantity captures the probability that the photon addition and subtraction processes are successfully realized through conditional measurements.

This distilled state is then used as an entanglement resource $\rho_{AB}$ for teleportation. Following distillation, teleportation is performed following the Braunstein-Kimble (BK) protocol. Alice prepares an unknown input coherent state $\ket{\alpha_{\text{in}}} = \hat{D}(\alpha_{\text{in}})\ket{0}$, where $\hat{D}(\alpha_{\text{in}})$ denotes the displacement operator, defined as follows:
\begin{equation}
   \hat{D}(\alpha) = \exp\left(\alpha \hat{a}^{\dagger} - \alpha^{*} \hat{a}\right).
\end{equation}
Then a balanced $50:50$ beam splitter is applied, yielding Eq. \eqref{Eq: beamsplitter} and \eqref{Eq: beamsplitter2}.
\begin{equation} \label{Eq: beamsplitter}
q_- = \frac{q_{\mathrm{in}} - q_A}{\sqrt{2}},
\end{equation}
and 
\begin{equation} \label{Eq: beamsplitter2}
    p_+ = \frac{p_{\mathrm{in}} + p_A}{\sqrt{2}}.
\end{equation}
\begin{figure}[tb]
    \centering
    \includegraphics[width=1\linewidth]{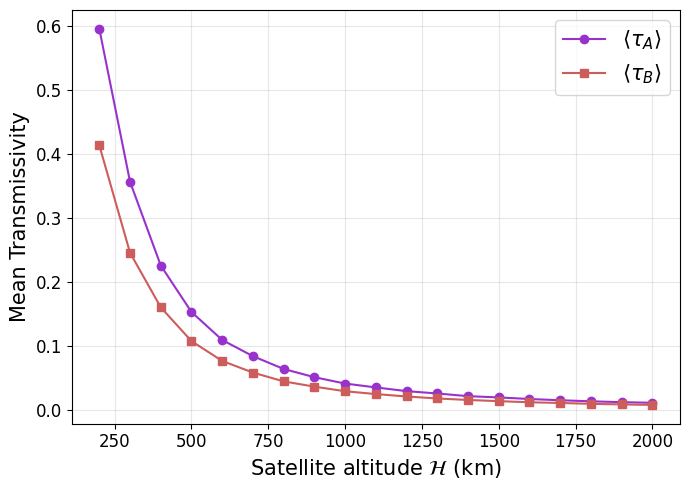}
    \caption{Average transmissivity $\langle \tau \rangle$ with respect to satellite altitude $\mathcal{H}$ for the dual downlink FSO channel.}
    \label{fig:tauab}
\end{figure}
Alice performs a homodyne measurement on $q_-$ and $ p _{+} $, obtaining classical values $q_m$ and $p_m$, which are transmitted to Bob through a classical channel. Then, Bob applies a displacement operation $\hat{D}(\alpha_m)$ with $\alpha_m = g(q_m + i p_m)$. In quadrature form, this corresponds to $\hat{q}_B \rightarrow \hat{q}_B + g \,\hat{q}_m $ and $\hat{p}_B \rightarrow \hat{p}_B + g \,\hat{p}_m $. After this correction, Bob's mode obtains the teleported output mode $\hat{a}_{\text{out}}$, which can be expressed as follows:
\begin{equation}
\hat{q}_{\text{out}} = \hat{q}_{\text{in}}+\hat{q}_N,
\end{equation}
and 
\begin{equation}
    \hat{p}_{\text{out}} = \hat{p}_{\text{in}}+\hat{p}_N,
\end{equation}
where $\hat{q}_N = \hat{q}_B - \hat{q}_A$ and $\hat{p}_N = \hat{p}_B + \hat{p}_A$, respectively. In the CF formalism, the output state is written as shown below. 
\begin{equation}
\begin{aligned}
    \chi_{\mathrm{out}}(-\zeta; \tau_A, \tau_B)
=& \chi_{\mathrm{in}}(-g \eta \, \zeta)\,
\chi_{AB}(-\zeta, -g \eta \, \zeta^*; \tau_A, \tau_B)
\\& \exp\left[-\frac{|\zeta|^2}{2} g^2 (1 - \eta^2)\right],
\end{aligned}
\end{equation}
where $\chi_{\mathrm{in}}(\zeta)$ denotes the CF of the input state, $\chi_{AB} (\zeta_A, \zeta_B)$ represent CF of the distilled resource, $g$ denotes the gain parameter, and $\eta^2$ is the homodyne detection efficiency. Finally, teleportation fidelity is given by the following equation:
\begin{equation}
    F (\tau_A, \tau_B) = \frac{1}{\pi} \int d^2 \zeta \, \chi_{\mathrm{in}}(\zeta)\, \chi_{\mathrm{out}}(-\zeta; \tau_A ,\tau_B).
\end{equation}
The mean fidelity is then obtained by the following:
\begin{equation}
\langle F \rangle = \int \int P(\tau_A)\, P(\tau_B)\, F(\tau_A, \tau_B)\, d\tau_A\, d\tau_B.
\end{equation}

\begin{figure} [t]
    \centering
    \includegraphics[width=1\linewidth]{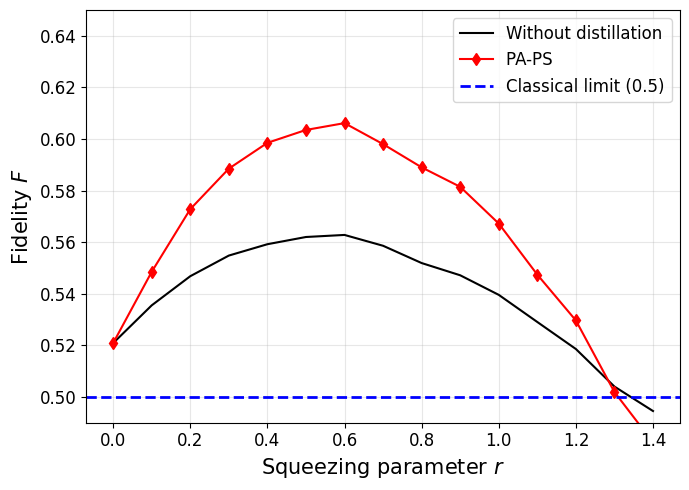}
    \caption{Fidelity of the proposed PA-PS as compared to FSO TMSV with respect to squeezing parameter $r$}
    \label{fig:fidelity_r}
\end{figure}

The entanglement of the shared TMSV state is quantified by negativity, which is obtained from the partial transpose of the density operator with respect to one mode. For state $\rho_{AB}$, negativity is defined as follows:
\begin{equation}
    N(\rho_{AB}) = \sum_{\mu_i < 0} |\mu_i|,
\end{equation}
where $\mu_i$ are the eigenvalues of the partially transposed density matrix. The corresponding logarithmic negativity can be defined as shown below.
\begin{equation}
    E_N(\rho_{AB}) = \log_2 \left( 1 + 2N(\rho_{AB}) \right).
\end{equation}
Before distillation, the transmitted TMSV state remains Gaussian and $E_N$ can be evaluated from the smallest symplectic eigenvalue $\tilde{\nu}_{-}$ of the partially transposed covariance matrix. In this case, the logarithmic negativity is given by the following:
\begin{equation}
    E_N = \max \left\{ 0,-\log_2 \left( 2\tilde{\nu}_{-} \right) \right\}.
\end{equation}
After applying the non-Gaussian PA-PS operation, the output state becomes non-Gaussian. Therefore, negativity is evaluated directly from the eigenvalues of the partially transposed density matrix in the truncated Fock basis.

\section{Numerical Results}\label{numerical_result}
In this section, we describe and analyze the performance of dual downlink PA-PS. The simulation parameter is listed in Table~\ref{tab: param}. In this simulation, the proposed PA-PS scheme was compared with noisy TMSV, which shares an identical FSO channel model structure and network design.

\begin{figure}[tb]
    \centering
    \includegraphics[width=1\linewidth]{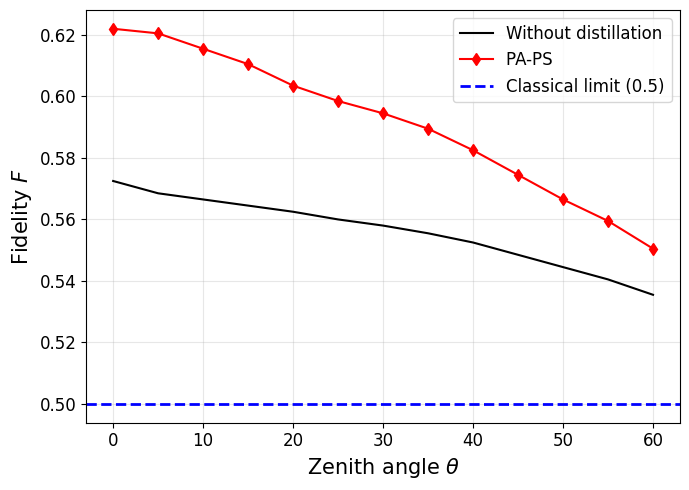}
    \caption{Fidelity $F$ with respect to zenith angle $\theta$ for dual-downlink FSO channel.}
    \label{fig: zenith}
\end{figure}

Figure~\ref{fig:tauab} shows the average channel transmissivity as a function of satellite altitude $\mathcal{H}$ for both links. As can be seen in Fig.~\ref{fig:tauab}, with increasing altitude, both $\langle \tau_A \rangle$ and $\langle \tau_B \rangle$ monotonically decreased, which can be attributed to diffraction loss and atmospheric attenuation. At low altitude, $\langle \tau_A \rangle$ was approximately $0.41$. As the altitude increased to around $500$~km, the transmissivities significantly decreased to approximately $0.15$ and $0.11$ for links $A$ and $B$. At higher altitude, particularly above $1000$~km, both links experienced severe attenuation, $\langle \tau_A \rangle$ and $\langle \tau_B \rangle$ decreasing below approximately $0.04$ and $0.03$, respectively. Throughout the entire altitude range, link $B$ consistently exhibited lower transmissivity, confirming that it represents the worst mode in the dual downlink configuration. Consequently, since the lowest transmissivity occurred in mode $B$, the PA–PS distillation was applied to mode $B$ in all subsequent simulations.

Figure~\ref{fig:fidelity_r} shows fidelity as a function of the squeezing parameter $r$. In this simulation, the satellite altitude was set to $\mathcal{H}=500$~km. Without distillation, fidelity reached a maximum of approximately $0.563$ at $r\approx 0.6$ and then gradually decreased due to increased sensitivity to channel noise at higher squeezing. With PA-PS, fidelity increased significantly, achieving approximately $0.606$ at $r \approx 0.6$. This represents an improvement of about $7.7\%$ over the non-distilled case. The results confirm that the existence of an optimal squeezing level, beyond which additional squeezing becomes detrimental due to loss-induced noise amplification.

Figure~\ref{fig: zenith} shows fidelity $F$ as a function of zenith angle $\theta$ for both cases with and without PA-PS distillation. In this simulation, the satellite altitude and squeezing parameter were set to $\mathcal{H}=500$~km and $r=0.6$, respectively. As shown in Fig.~\ref{fig: zenith}, fidelity in both cases exhibited an overall decreasing trend as the zenith angle increased. At small zenith angles $(\theta \approx0^{\circ})$, fidelity without distillation was approximately $0.57$. In comparison, the PA–PS-assisted case achieved a higher value of approximately $0.62$, demonstrating a noticeable performance improvement due to the distillation process. As the zenith angle increased to approximately $30^\circ$, both fidelities decreased to $0.535$ and $0.555$ for the cases without distillation and PA–PS, respectively. At larger zenith angles ($\theta>50^\circ$), the fidelity decreased further, reaching approximately $0.53$ without distillation and $0.535$ with PA–PS. Throughout the entire range of zenith angles, the PA–PS distillation consistently provided higher fidelity compared to the without distillation case, confirming its effectiveness in enhancing shared entanglement under varying propagation conditions. Moreover, both curves remained above the classical limit ($F=0.5$), indicating that quantum teleportation was successfully maintained for all considered zenith angles. 

\begin{figure} [t]
    \centering
    \includegraphics[width=1\linewidth]{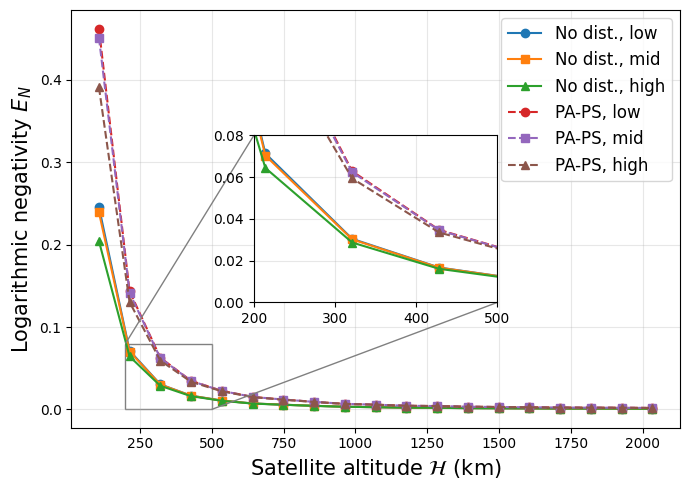}
    \caption{Logarithmic negativity $E_N$ with respect to satellite altitude $\mathcal{H}$ for dual downlink FSO channels under low, mid, and high turbulence conditions.}
    \label{fig: negativity_turb}
\end{figure}

Figure~\ref{fig: negativity_turb} shows the logarithmic negativity $E_N$ as a function of satellite altitude $\mathcal{H}$ under low, mid, and high turbulence conditions for both cases without distillation and PA-PS distillation. This channel behavior can be directly explained by the role of the HV turbulence parameters $A_{\mathrm{HV}}$ and $v$. The turbulence parameters were defined as follows: low $(A_{\mathrm{HV}} = 10^{-15}, v = 5 \, \text{ms})$, mid $(A_{\mathrm{HV}} = 2.75 \times 10^{-14}, v = 21 \, \text{ms})$, and high $(A_{\mathrm{HV}} = 10^{-13}, v = 50 \,\text{ms})$. 

As shown in Fig.~\ref{fig: negativity_turb}, negativity curves corresponding to different turbulence levels were closely overlapped throughout the satellite altitude range, indicating that the impact of turbulence was not significant in the considered scenario. This behavior can be attributed to the fact that turbulence primarily affected the beam through spreading and wandering, which introduced only minor fluctuations in the effective transmissivity. By contrast, the dominant degradation mechanism was propagation loss due to diffraction and atmospheric attenuation, which caused a rapid decay of negativity with increasing altitude. Consequently, the variations in $A_{\mathrm{HV}}$ and $v$ led to only small perturbations in the average channel conditions, resulting in nearly identical performance in all turbulence regimes. These results indicated that, under weak-to-moderate turbulence conditions, system performance was primarily limited by satellite altitude rather than turbulence strength.

\begin{figure} [t]
    \centering
    \includegraphics[width=1\linewidth]{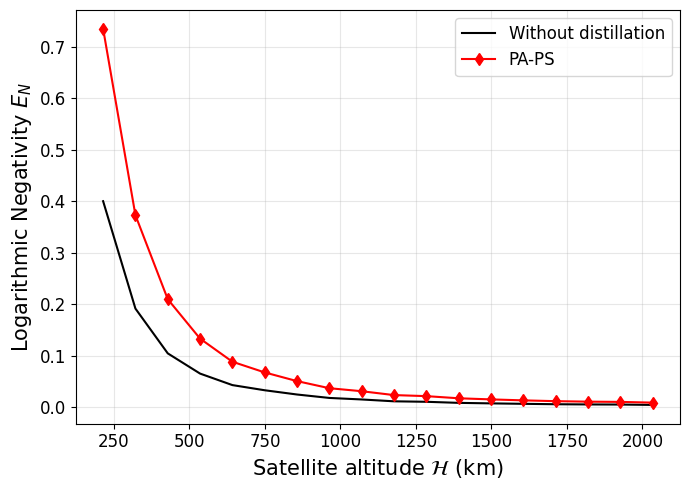}
    \caption{Logarithmic negativity $E_N$ of PA-PS as compared to satellite altitude $\mathcal{H}$ for dual-downlink FSO channel.}
    \label{fig: negativity}
\end{figure}

Figure~\ref{fig: negativity} shows the logarithmic negativity as a function of satellite altitude $\mathcal{H}$ for the cases with and without PA-PS distillation. Without distillation, logarithmic negativity $E_N$ rapidly decreases from approximately $0.400$ at $200$~km to nearly $0.004$ at $2000$~km, indicating severe degradation of entanglement caused by increasing channel attenuation. By contrast, the PA–PS protocol significantly improved logarithmic negativity, achieving an initial value of approximately $E_N \approx 0.738$ at $200$~km, which was much higher than the case without distillation. At intermediate altitude (e.g., $500$~km), the logarithmic negativity remained approximately $E_N \approx 0.133$ with PA-PS, compared to $E_N = 0.065$ without distillation. This corresponds to an improvement of about $105\%$, indicating that the entanglement is more than doubled. However, similarly to fidelity behavior, this advantage gradually diminished at longer distances, where both converged to zero entanglement due to extreme attenuation in the FSO channel.

\begin{figure} [t]
    \centering
    \includegraphics[width=1\linewidth]{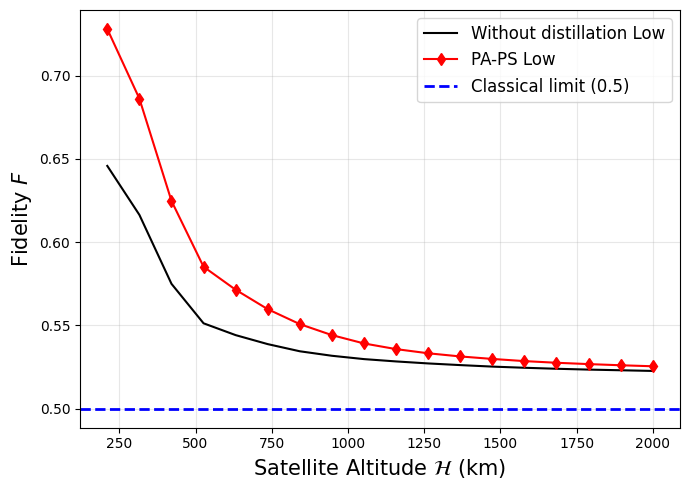}
    \caption{Fidelity $F$ with respect to satellite altitude $\mathcal{H}$ for dual-downlink FSO channel.}
    \label{fig: fidelity}
\end{figure}

Figure~\ref{fig: fidelity} shows the performance of the proposed PA-PS distillation protocol in terms of fidelity $F$ with respect to satellite altitude $\mathcal{H}$. In this simulation, the squeezing parameter was set to $r=0.6$. The results show that fidelity decreased as the satellite altitude increased because of the stronger propagation loss in the dual downlink channels. Without distillation, fidelity decreased from approximately $0.646$ at $200$~km to $0.522$ at $2000$~km. By contrast, the PA-PS scheme consistently achieved higher fidelity, decreasing from approximately $0.768$ to $0.537$ over the same altitude range.  The improvement was most significant in the short-to-medium altitude ($0–600$~km), where the fidelity gain amounted to about $0.03–0.09$, indicating that PA–PS can compensate for moderate channel loss. However, at long distances ($>1500$~km), both curves converge towards the classical limit of $0.5$, showing that extreme loss limits the effectiveness of distillation.

\begin{figure} [t]
    \centering
    \includegraphics[width=1\linewidth]{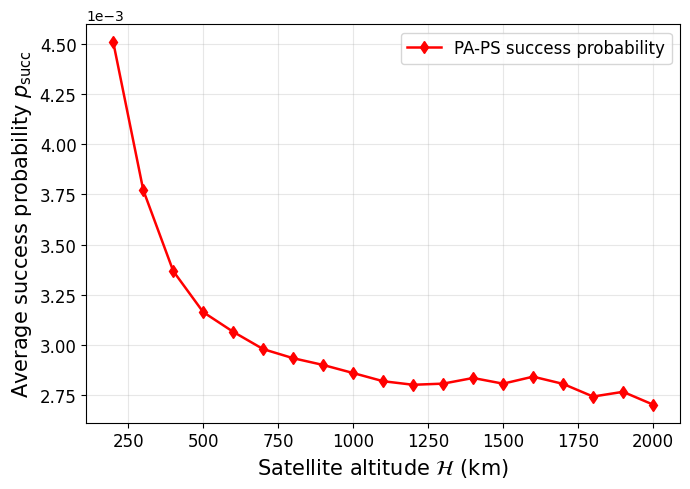}
    \caption{Average success probability of the PA-PS distillation with respect to satellite altitude $\mathcal{H}$.}
    \label{fig: p_succ}
\end{figure}

Figure~\ref{fig: p_succ} shows the average success probability $p_{\mathrm{succ}}$ of the PA-PS distillation as a function of satellite altitude $\mathcal{H}$. In this simulation, the squeezing parameter was set to $r = 0.6$. As shown in Fig.~\ref{fig: p_succ}, the success probability exhibited an overall decreasing trend as the satellite altitude increased. At $\mathcal{H}=200$~km the success probability was approximately $4.51\times10^{-3}$ (i.e., $0.45\%$), and decreased rapidly to about $3.16\times10^{-3}$($\sim0.32\%$) at $\mathcal{H}=500$~km. In the intermediate altitude ($500-1000$~km), the success probability decreased more gradually, reaching approximately $2.86\times10^{-3}$ $(\sim 0.29\%)$ at $\mathcal{H}=1000$~km. Beyond this range, $p_{\mathrm{succ}}$ showed minor fluctuations around a nearly constant value, stabilizing around $2.70-2.84 \times10^{-3}$ (i.e., $0.27\%-0.28\%$) for $\mathcal{H}>1000$~km.

This trend was well explained by the analytical scaling of the PA-PS success probability. For a TMSV state, the mean photon number was given by $\langle n \rangle = \sinh^{2}{(r)}$, which yielded $\langle n \rangle \approx 0.40$ for $r=0.6$. The success probability of a single photon subtraction event scaled as $p_{\mathrm{PS}} \sim (1-T_{\mathrm{S}})\langle n \rangle$, while the sequential PA-PS operation required two heralding events, leading to an overall scaling of $p_{\mathrm{succ}} \sim (1-T_{\mathrm{S}})^2\langle n \rangle$. For $T_{\mathrm{S}}=0.95$, this gave $p_{\mathrm{succ}}\sim (0.05)^2\times0.40\approx1.0\times10^{-3}$ (i.e., $\sim0.10\%$). The factor $\sim3-4$ difference between this estimate and the simulated values ($\sim 2.7-4.5 \times10^{-3}$, or $0.27\% - 0.45\%$) was attributed to multi-photon contributions and higher-order terms not captured in the single-photon approximation. The higher values observed at shorter distances arose from 
better preserved quantum states due to lower channel loss, whereas the reduction and eventual convergence at higher altitudes were primarily due to increased photon loss, which suppressed the probability of successful heralding events.



Moreover, this figure highlights the inherent trade-off between fidelity and success probability. While the PA-PS distillation improves the fidelity of the shared entangled state, this enhancement comes at the cost of a reduced success probability, particularly at longer distances. As a result, although fidelity can still be increased through distillation, the likelihood of successful operation becomes increasingly limited.

\section{Conclusion}\label{conclusion}
In this paper, we investigated the performance of CV-QT over a dual downlink FSO channel. The results revealed that channel transmissivity significantly decreased with distance, leading to degradation in both entanglement and fidelity. To mitigate this degradation, we applied a non-Gaussian distillation protocol based on the sequential combination of PA-PS on the worst channel. Our findings showed that the proposed scheme consistently enhanced system performance, yielding up to approximately $7.7~\%$ improvement in fidelity and about $105~\%$ enhancement in entanglement negativity in the low-to-moderate regime. Moreover, an optimal squeezing parameter was identified, beyond which performance degraded due to increased sensitivity to channel noise and loss.

Despite these improvements, several limitations remain. First, the PA-PS protocol is inherently probabilistic, and its success probability introduces a non-negligible resource cost, especially in high-loss regimes. Second, the performance gain diminishes at long distances (e.g., beyond $\sim\!1500$~km), where severe attenuation dominates and distillation becomes less effective. These factors highlight a trade-off between performance enhancement and operational feasibility.

Future work can address these challenges by exploring more advanced strategies, such as multi-photon PA–PS operations to further enhance non-Gaussianity, adaptive squeezing optimization based on real-time channel conditions, and extending the framework to uplink scenarios, where turbulence effects differ significantly. Additionally, integrating the proposed scheme into multi-hop quantum networks with entanglement swapping could provide a scalable solution for long-distance quantum communication.

\begin{thebibliography}{99}

\bibitem{bennett1993teleporting}
C. H. Bennett, G. Brassard, C. Cr\'epeau, R. Jozsa, A. Peres, and W. K. Wootters,
``Teleporting an unknown quantum state via dual classical and Einstein--Podolsky--Rosen channels,''
\emph{Physical Review Letters}, vol. 70, no. 13, pp. 1895--1899, 1993.

\bibitem{cacciapuoti2020QT}
A. S. Cacciapuoti, M. Caleffi, R. Van Meter, and L. Hanzo,
``When entanglement meets classical communications: Quantum teleportation for the quantum internet,''
\emph{IEEE Trans. Commun.}, vol. 68, no. 6, pp. 3808--3833, Jun. 2020.

\bibitem{chehimi2022physics}
M. Chehimi and W. Saad,
``Physics-informed quantum communication networks: A vision toward the quantum internet,''
\emph{IEEE Netw.}, vol. 36, no. 5, pp. 32--38, Oct. 2022.

\bibitem{CV_QC}
V. C. Usenko, A. Ac\'in, R. All\'eaume, U. L. Andersen, E. Diamanti, T. Gehring, A. A. E. Hajomer, F. Kanitschar, C. Pacher, S. Pirandola, and V. Pruneri,
``Continuous-variable quantum communication,''
\emph{Reviews of Modern Physics}, vol. 98, no. 1, Art. no. 015003, Mar. 2026.

\bibitem{shi2023continuous}
S. Shi, Y. Wang, L. Tian, W. Li, Y. Wu, Q. Wang, Y. Zheng, and K. Peng,
``Continuous variable quantum teleportation network,''
\emph{Laser \& Photonics Reviews}, vol. 17, no. 2, Art. no. 2200508, 2023.

\bibitem{nguyen2023enhancing}
T. V. Nguyen, H. T. Le, H. T. T. Pham, V. Mai, and N. T. Dang,
``Enhancing design and performance analysis of satellite entanglement-based CV-QKD/FSO systems,''
\emph{IEEE Access}, vol. 11, pp. 112097--112107, 2023.

\bibitem{fiber_bipartite}
W. Dai, T. Peng, and M. Z. Win,
``Optimal Remote Entanglement Distribution,''
\emph{IEEE J. Sel. Areas in Commun.}, vol. 38, no. 3, pp. 540--556, Mar. 2020.

\bibitem{survey_cv_satellite}
N. Hosseinidehaj, Z. Babar, R. Malaney, S. X. Ng, and L. Hanzo,
``Satellite-based continuous-variable quantum communications: State-of-the-art and a predictive outlook,''
\emph{IEEE Commun. Surveys \& Tutorials}, vol. 21, no. 1, pp. 881--919, First Quarter 2019.

\bibitem{chehimi2025reconfigurable}
M. Chehimi, M. Elhattab, W. Saad, G. Vardoyan, N. K. Panigrahy, C. Assi, and D. Towsley,
``Reconfigurable intelligent surface (RIS)-assisted entanglement distribution in FSO quantum networks,''
\emph{IEEE Trans. Wireless Commun.}, vol. 24, no. 4, pp. 3132--3148, Apr. 2025.

\bibitem{pirandola_cv-sat}
T. Gonzalez-Raya, S. Pirandola, and M. Sanz,
``Satellite-based entanglement distribution and quantum teleportation with continuous variables,''
\emph{Commun. Physics}, vol. 7, no. 1, Art. no. 126, 2024.

\bibitem{uplink_cv}
Z. Wang, T. C. Ralph, R. Aguinaldo, and R. Malaney,
``Exploiting Spatial Diversity in Earth-to-Satellite Quantum-Classical Communications,''
\emph{IEEE Trans. on Commun.}, vol. 73, no. 11, pp. 11736--11752, 2025.

\bibitem{lee2013distillation}
J. Lee and H. Nha,
``Entanglement distillation for continuous variables in a thermal environment: Effectiveness of a non-Gaussian operation,''
\emph{Physical Review A}, vol. 87, no. 3, Art. no. 032307, 2013.

\bibitem{xiang2010NLA}
G.-Y. Xiang, T. C. Ralph, A. P. Lund, N. Walk, and G. J. Pryde,
``Heralded noiseless linear amplification and distillation of entanglement,''
\emph{Nature Photonics}, vol. 4, no. 5, pp. 316--319, 2010.

\bibitem{xia2025_QS}
Y. Xia, Y. Wang, T. Xiao, W. Ye, Z. Liao, and X. Zhou,
``Enhancing quantum entanglement with the local displacement-based quantum scissor,''
\emph{Chinese Journal of Physics}, vol. 95, pp. 731--741, 2025.

\bibitem{zhang2010photonsubstraction}
S. Li Zhang and P. van Loock,
``Distillation of mixed-state continuous-variable entanglement by photon subtraction,''
\emph{Physical Review A}, vol. 82, no. 6, Art. no. 062316, 2010.

\bibitem{wang2024PSswapping}
N. Wang, M. Wang, and X. Su,
``Continuous-variable entanglement swapping with photon subtraction,''
\emph{Physical Review A}, vol. 110, no. 1, Art. no. 012416, 2024.

\bibitem{lee2011psa}
S.-Y. Lee, S.-W. Ji, H.-J. Kim, and H. Nha,
``Enhancing quantum entanglement for continuous variables by a coherent superposition of photon subtraction and addition,''
\emph{Physical Review A}, vol. 84, no. 1, Art. no. 012302, 2011.

\bibitem{villasenor2021distillnon}
E. Villase\~nor and R. Malaney,
``Enhancing continuous variable quantum teleportation using non-Gaussian resources,''
in \emph{Proc. IEEE Global Commun. Conference (GLOBECOM)}, Madrid, Spain, 2021, pp. 1--6.

\bibitem{CV_book}
A. Serafini,
\emph{Quantum Continuous Variables: A Primer of Theoretical Methods}.
Boca Raton, FL, USA: CRC Press, 2023.

\bibitem{hufnagel1964modulation}
R. E. Hufnagel and N. R. Stanley,
``Modulation transfer function associated with image transmission through turbulent media,''
\emph{Journal of the Optical Society of America}, vol. 54, no. 1, pp. 52--61, 1964.

\end{thebibliography}


\end{document}